\documentclass[conference]{IEEEtran}
\usepackage{amsmath,amsfonts}
\usepackage{algorithmic}
\usepackage{algorithm}
\usepackage{array}
\usepackage[caption=false,font=normalsize,labelfont=sf,textfont=sf]{subfig}
\usepackage{textcomp}
\usepackage{stfloats}
\usepackage{url}
\usepackage{verbatim}
\usepackage{graphicx}
\usepackage{cite}
\usepackage{xcolor}
\usepackage{mathrsfs}

\title{HAPS-RIS or HAPS-Relay: Which Outperforms Under Impairments with NOMA in 6G NTN?}
\author{
\IEEEauthorblockN{Bilal Karaman\IEEEauthorrefmark{1}, Faicel Khennoufa\IEEEauthorrefmark{2}, Ilhan Basturk\IEEEauthorrefmark{1}, Metin Ozturk\IEEEauthorrefmark{3},\\ 
Ferdi Kara\IEEEauthorrefmark{4}, Sezai Taskin\IEEEauthorrefmark{1}, Halim Yanikomeroglu\IEEEauthorrefmark{5}}
\IEEEauthorblockA{\IEEEauthorrefmark{1}Manisa Celal Bayar University, Manisa, Türkiye}
\IEEEauthorblockA{\IEEEauthorrefmark{2}National Higher School of Advanced Technologies (ENSTA), Algiers, Algeria}
\IEEEauthorblockA{\IEEEauthorrefmark{3}Ankara Yıldırım Beyazıt University, Ankara, Türkiye}
\IEEEauthorblockA{\IEEEauthorrefmark{4}Ericsson Research, Lund, Sweden}
\IEEEauthorblockA{\IEEEauthorrefmark{5}Non-Terrestrial Networks Lab (Carleton-NTN), Carleton University, Ottawa, ON, Canada}
}

\begin{document}




\maketitle

\begin{abstract}

This paper investigates the performance of high-altitude platform station (HAPS)-assisted communication systems employing either reconfigurable intelligent surfaces (RIS) or relay stations (RS) under non-orthogonal multiple access (NOMA) scheme. Practical system impairments, including hardware impairments (HWI) and imperfect channel state information (CSI), are explicitly considered. The results show that HAPS-RIS outperforms HAPS-RS in terms of both sum-rate and energy efficiency under non-ideal conditions due to its passive nature, which avoids noise amplification. Furthermore, it is demonstrated that RIS element allocation and user spatial distribution significantly impact NOMA performance, where increased user separation and proper allocation enhance channel disparity and improve system efficiency. 
\color{black}
Despite its higher sensitivity to imperfect CSI, HAPS-RIS can effectively compensate for performance degradation through large-scale RIS element deployment, maintaining a performance advantage over half-duplex RS-based systems. 
\color{black}
These insights provide useful design guidelines for impairment-aware HAPS-assisted 6G communication systems.

\end{abstract}

\begin{IEEEkeywords}
High-altitude platform station (HAPS), non-terrestrial networks (NTN), reconfigurable intelligent surfaces (RIS), relay station (RS), HWI, imperfect CSI.
\end{IEEEkeywords}

\section{Introduction}
The stringent requirements of sixth-generation (6G) wireless systems, including immersive communication, massive connectivity, and hyper-reliable low-latency services, while ensuring ubiquitous coverage and sustainability, as outlined in the IMT-2030 framework~\cite{imt2030Report}, make sole reliance on terrestrial networks increasingly impractical.
As a result, non-terrestrial networks~(NTN) have emerged as a key enabler to complement existing infrastructures and provide seamless global connectivity~\cite{ccilouglu2025strategic}.
Among NTN solutions, high-altitude platform stations (HAPS) have gained significant attention due to their unique characteristics. Operating at altitudes of around 20 km, HAPS offer a favorable trade-off between coverage, latency, and deployment flexibility~\cite{karaman2025solutions}. Compared to low-Earth orbit (LEO) satellites, HAPS experience lower path loss and reduced latency, while providing wider coverage than uncrewed aerial vehicles (UAVs). 
HAPS can be rapidly deployed to support emergency scenarios, temporary events, and rural connectivity, and their large payload capacity allows the integration of advanced communication technologies. Furthermore, thanks to photovoltaic~(PV) panels and onboard energy storage systems, HAPS can sustain long-duration operations, making them a promising candidate for sustainable 6G architectures. 

From a communication payload perspective, HAPS can be equipped with different technologies, including multi-antenna base stations (BSs), relay stations (RSs), and reconfigurable intelligent surfaces (RIS). While HAPS-mounted BS solutions can provide high-capacity communication and advanced processing capabilities, they incur substantial power consumption, typically in the order of several kilowatts (e.g., 6--9 kW) due to radio frequency (RF) chains, signal processing units, and hardware requirements~\cite{CHENG2022121672}. Alternatively, HAPS-relay station (HAPS-RS), which operate based on amplify-and-forward or decode-and-forward principles, offer a lighter architecture but still consume considerable power, typically on the order of 1 kW, due to active transmission and signal processing~\cite{alfattani2022multi}. In contrast, RIS technology introduces a nearly passive communication paradigm by enabling intelligent signal reflection without requiring dedicated RF chains. Owing to their low power consumption and ease of deployment, RIS are particularly attractive for energy-constrained platforms such as HAPS~\cite{karamanIoT}. Hence, the integration of RIS with HAPS (i.e., HAPS-RIS) has emerged as a promising solution to enhance coverage and improve spectral and energy efficiency~\cite{karaman2025trade}.

Motivated by these advantages, recent studies have investigated HAPS-RIS-assisted communication systems in various contexts. Prior works have explored sum-rate maximization, reflecting element allocation, and resource efficiency optimization for beyond-cell communications supported by HAPS-RIS architectures~\cite{safwanGlobecom, safwanLetter}. In addition, HAPS-RIS systems have been considered for post-disaster communications~\cite{matracia2024unleashing}, backhaul connectivity~\cite{karamanIoT}, hybrid aerial-terrestrial deployments, and integrated communication scenarios~\cite{azizi2025exploring}. Furthermore, comparative analyses between HAPS-RIS and conventional HAPS-RS have demonstrated that RIS-assisted solutions can achieve superior energy efficiency and competitive performance under certain conditions~\cite{alfattani2022multi}. However, despite these efforts, a comprehensive comparison of HAPS-RIS and HAPS-RS architectures under realistic system impairments and advanced multiple access schemes is still lacking. 

In this context, the impact of hardware impairments (HWI) and imperfect channel state information (CSI), which are inevitable in practical HAPS deployments, has not been thoroughly investigated for HAPS-RIS and HAPS-RS systems. \color{black} While HAPS-RS suffers from half-duplex operation, HAPS-RIS-assisted systems can simultaneously serve multiple users via element allocation. 
On the other hand, RIS is more sensitive to imperfect CSI, and we hypothesize in this study that this degradation can be effectively mitigated by \color{black}a large number of RIS elements, \color{black} leading to a non-trivial performance trade-off.
\color{black} Moreover, the integration of non-orthogonal multiple access~(NOMA) with HAPS-assisted architectures remains largely unexplored, despite its potential to enhance spectral efficiency and user connectivity. Existing studies also do not address how RIS elements allocation and user pairing interact under NOMA-based transmission, nor do they consider the effect of inter-user distance on system performance.

To address these gaps, this paper develops a unified analytical framework to systematically evaluate and compare HAPS-RIS and HAPS-RS-assisted communication systems under both orthogonal multiple access~(OMA) and NOMA transmission schemes, while explicitly accounting for practical system impairments. 
\textcolor{black}{A distinguishing aspect of this work is the joint consideration of HAPS-RIS and HAPS-RS architectures under HWI and imperfect CSI conditions, together with a systematic characterization of their spectral- and energy-efficiency trade-offs.}
Furthermore, this work provides design insights into RIS elements allocation in NOMA systems and demonstrates the critical role of inter-user distance in user pairing strategies. 

The main contributions of this work are as follows:
\begin{itemize}
    \item A unified system model for HAPS-RIS and HAPS-RS under a NOMA scheme is proposed.
    \item We incorporate HWI and imperfect CSI into the analysis and evaluate their impact on system performance.
    \item We reveal the fundamental spectral and energy efficiency trade-offs between HAPS-RIS and HAPS-RS architectures under realistic conditions.
    \item RIS element allocation strategies in NOMA systems are studied, and their effect on performance is quantified.
    \item We analyze the impact of inter-user distance on user pairing in NOMA-based HAPS systems.
\end{itemize}


\section{System Model}

As illustrated in Fig.~1, we consider a HAPS-assisted communication system consisting of a control station~(CS), a HAPS platform equipped with either an RS (i.e., HAPS-RS) or a RIS (i.e., HAPS-RIS) with $N$ elements, and multiple ground user equipments (UEs). The CS is equipped with a high-gain antenna, while all UEs are assumed to be single-antenna devices.
The CS maintains a strong line-of-sight (LoS) link with the HAPS; however, there is no direct link between the CS and the UEs. All communications are realized via the HAPS. \color{black} In the HAPS-RS case, the HAPS operates as a half-duplex active relay, while in the HAPS-RIS case, it passively reflects the incident signals toward the UEs without RF chains. \color{black} We also take into account practical limitations, including HWI and imperfect CSI. The channels are modeled by incorporating estimation errors, and the impact of HWI is included at both the CS and the UEs. 

\begin{figure}[!t]
\centering
\includegraphics[width=2.7in]{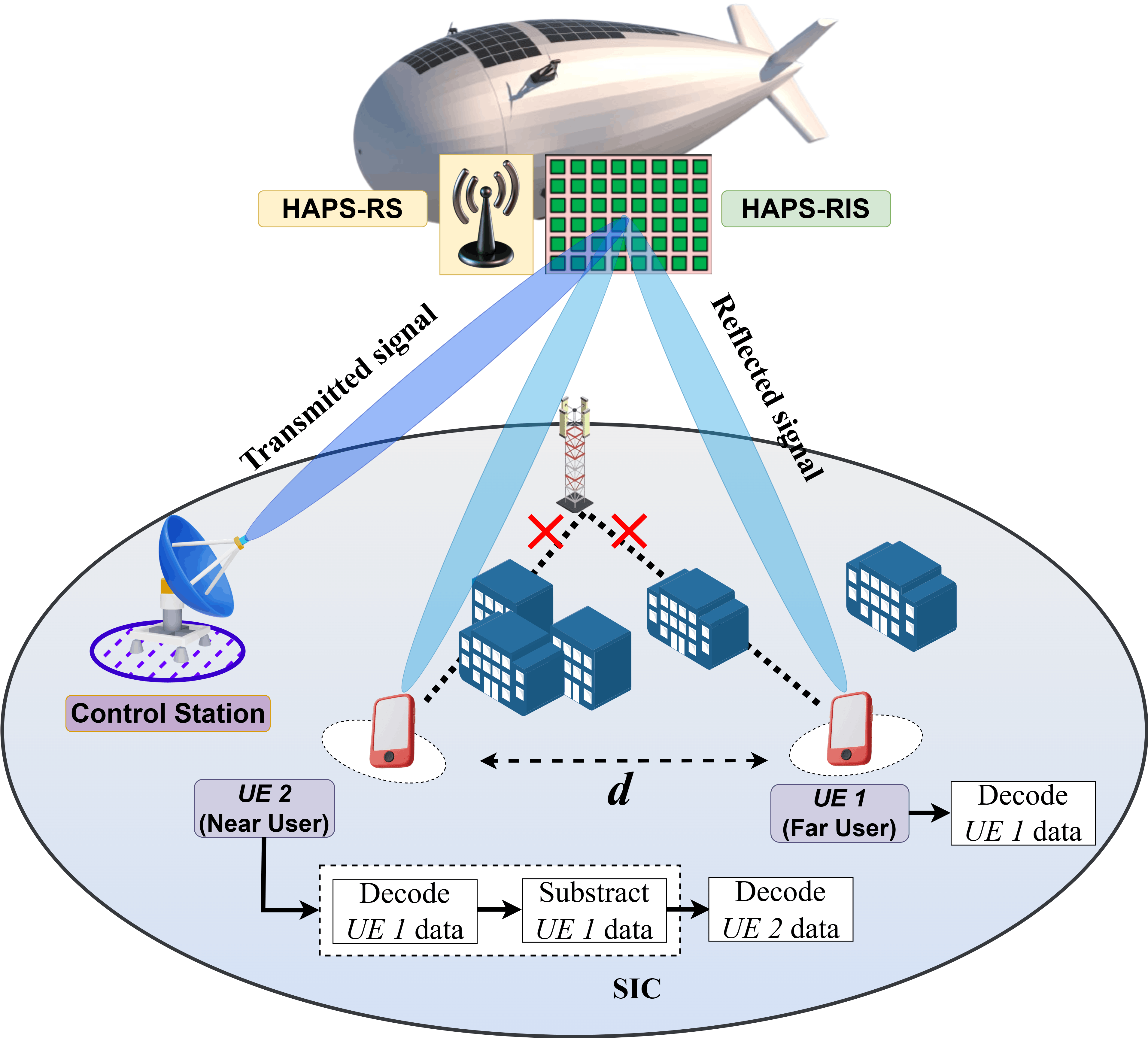}
\caption{System model of the HAPS-assisted communication framework with RIS- and RS-based architectures under NOMA transmission.}
\end{figure}

\subsection{HAPS-RS-assisted Communication}

The CS transmits a superposition coding signal, where the power allocation coefficients are assigned according to the ordered effective channel gains of the indirect links, i.e., 
$\left| h^{\text{RS}}_{\text{ch}}h^{\text{RS}}_{\text{hu,1}} \right|^{2} 
< \left| h^{\text{RS}}_{\text{ch}}h^{\text{RS}}_{\text{hu,2}} \right|^{2} 
< ... < \left| h^{\text{RS}}_{\text{ch}}h^{\text{RS}}_{\text{hu},i} \right|^{2}$. \color{black}Here, $h^{\text{RS}}_{\text{ch}}$ denotes the CS-to-HAPS-RS channel, $h^{\text{RS}}_{\text{hu},i}$ represents the HAPS-RS–to–$i$th user channel. \color{black}
Accordingly, the received signal at the $i$-th ground user via HAPS-RS is expressed as
\begin{equation}
\begin{split}
y_{\text{RS},i} &= h^{\text{RS}}_{\text{hu},i} \Big( \Lambda \sqrt{P^{\text{RS}}_{\text{t}}} \Big( 
h^{\text{RS}}_{\text{ch}} \big( \sqrt{P^{\text{CS}}_{\text{t}}}x_{\text{sc}} + \eta_{\text{t,ch}} \big) \\
&\quad + \eta_{\text{r,ch}} + \aleph \Big) + \eta_{\text{t,hu},i} \Big) 
+ \eta_{\text{r,hu},i} + \aleph,
\end{split}
\end{equation}
where $\Lambda = \sqrt{\frac{1}{\left| h^{\text{RS}}_{\text{ch}} \right|^{2}P^{\text{CS}}_{\text{t}}+\sigma^{2}}}$. The CS--HAPS-RS and HAPS-RS--user channels are assumed to be imperfectly estimated and are modeled independently as $h^{\text{RS}}_{\text{ch}} = \hat{h}^{\text{RS}}_{\text{ch}} + \Delta h^{\text{RS}}_{\text{ch}}$ and $h^{\text{RS}}_{\text{hu},i} = \hat{h}^{\text{RS}}_{\text{hu},i} + \Delta h^{\text{RS}}_{\text{hu},i}$,
where $\hat{h}^{\text{RS}}_{\text{ch}}$ and $\hat{h}^{\text{RS}}_{\text{hu},i}$ denote the estimated channel coefficients, while 
$\Delta h^{\text{RS}}_{\text{ch}}$ and $\Delta h^{\text{RS}}_{\text{hu},i}$ represent the corresponding estimation errors. 
The estimation errors are modeled as independent complex Gaussian random variables with zero mean and variance $\epsilon$, i.e., 
$\Delta h^{\text{RS}}_{\text{ch}} \sim \mathcal{CN}(0,\epsilon)$ and 
$\Delta h^{\text{RS}}_{\text{hu},i} \sim \mathcal{CN}(0,\epsilon)$, 
where $\epsilon$ denotes the level of CSI imperfection. 
The transmitted superposition signal is given by 
$x_{\text{sc}}=\sum_{i=1}^{L}\alpha_{i}s_{i}$, 
where \color{black} $L$ denotes the total number of users, \color{black} $\alpha_{i}$ denotes the power allocation coefficient satisfying 
$\alpha_{1} > \alpha_{2} > ... > \alpha_{L}$ and 
$\sum_{i=1}^{L} \alpha_i = 1$, and $s_i$ represents the signal intended for the $i$-th user. Moreover,
$P^{\text{CS}}_{\text{t}}$ and $P^{\text{RS}}_{\text{t}}$ correspond to the transmit powers of the CS and RS, respectively.

The noise term $\aleph$ follows a complex Gaussian distribution, i.e., 
$\aleph \sim \mathcal{CN}(0, \sigma^2)$. 
The noise variance is expressed as $\sigma^2 = k_\text{B} T_{\text{sys}} B N_\text{F}$, 
where $k_\text{B}$ denotes the Boltzmann constant, $T_{\text{sys}}$ is the system temperature, 
$B$ represents the system bandwidth, and $N_\text{F}$ corresponds to the receiver noise figure. The distortion noises $\eta_{\text{t,ch}}$, $\eta_{\text{r,ch}}$, $\eta_{\text{t,hu},i}$, and $\eta_{\text{r,hu},i}$ are modeled as independent Gaussian random variables~\cite{khennoufa2025multi}, with 
$\eta_{\text{t,ch}}\sim \mathcal{CN}(0, k^{2}_{\text{t,ch}}P^{\text{CS}}_{\text{t}})$, 
$\eta_{\text{r,ch}}\sim \mathcal{CN}(0, k^{2}_{\text{r,ch}}P^{\text{CS}}_{\text{t}}\left| h^{\text{RS}}_{\text{ch}} \right|^{2})$, 
$\eta_{\text{t,hu},i}\sim \mathcal{CN}(0, k^{2}_{\text{t,hu},i}P^{\text{RS}}_{\text{t}})$, and 
$\eta_{\text{r,hu},i}\sim \mathcal{CN}(0, k^{2}_{\text{r,hu},i}P^{\text{RS}}_{\text{t}}\left| h^{\text{RS}}_{\text{hu},i} \right|^{2})$. 
The parameters $k_{\text{t,ch}}$, $k_{\text{r,ch}}$, $k_{\text{t,hu},i}$, and $k_{\text{r,hu},i}$ quantify the impairment levels at the transmitter and receiver sides. 
The overall impact of transceiver impairments is characterized by 
$ K^{2}_{\text{ch}} = k^{2}_{\text{t,ch}} + k^{2}_{\text{r,ch}}$ and 
$ K^{2}_{\text{hu},i} = k^{2}_{\text{t,hu},i} + k^{2}_{\text{r,hu},i}$.

The channel between the CS and HAPS-RS follows a Rician distribution and is given by

\begin{equation}
    h^{\text{RS}}_{\text{ch}} = \sqrt{\frac{G_{\text{CS}}G_{\text{RS}}}{\zeta_{\text{ch}}}\frac{Z_{\text{ch}}}{Z_{\text{ch}}+1}}\bar{h}^{\text{RS}}_{\text{ch}} + \sqrt{\frac{G_{\text{CS}}G_{\text{RS}}}{\zeta_{\text{ch}}}\frac{1}{Z_{\text{ch}}+1}}\tilde{h}^{\text{RS}}_{\text{ch}},
\end{equation}
while the channel between the HAPS-RS and the $i$-th ground user is expressed as

\begin{equation}
    h^{\text{RS}}_{\text{hu},i} = \sqrt{\frac{G_{\text{RS}}G_{\text{U}}}{\zeta_{\text{hu},i}}\frac{Z_{\text{hu},i}}{Z_{\text{hu},i}+1}}\bar{h}^{\text{RS}}_{\text{hu},i} + \sqrt{\frac{G_{\text{RS}}G_{\text{U}}}{\zeta_{\text{hu},i}}\frac{1}{Z_{\text{hu},i}+1}}\tilde{h}^{\text{RS}}_{\text{hu},i},
\end{equation}
where $G_{\text{CS}}$, $G_{\text{RS}}$, and $G_{\text{U}}$ denote the antenna gains of the CS, RS, and users, respectively. $Z_{\text{ch}}$ and $Z_{\text{hu},i}$ represent the Rician factors, while $\zeta_{\text{ch}}$ and $\zeta_{\text{hu},i}$ correspond to the large-scale path-loss coefficients. The LoS and non-LoS~(NLoS) components are denoted by $\bar{h}$ and $\tilde{h}$, respectively, where the NLoS components follow $\mathcal{CN}(0,1)$.
Considering the presence of HWI and imperfect CSI, the signal-to-interference-plus-noise ratio (SINR) for the $i$-th user is expressed as

\begin{equation}
    \gamma_{\text{RS},s_{i}}= \frac{\alpha_{1}\left( \Lambda^{2}P^{\text{RS}}_{\text{t}}P^{\text{CS}}_{\text{t}} \left| h^{\text{RS}}_{\text{ch}} \right|^{2} \left| h^{\text{RS}}_{\text{hu},i} \right|^{2} \right)}{\mathscr{C}_{\text{a}} + \Delta_{\text{a}}+\left( 1+\left| h^{\text{RS}}_{\text{hu},i} \right|^{2}\Lambda^{2}P^{\text{RS}}_{\text{t}} \right)\sigma^{2}},
\end{equation}
where $\Delta_{\text{a}} = \left| h^{\text{RS}}_{\text{ch}} \right|^{2}\left| h^{\text{RS}}_{\text{hu},i} \right|^{2}\Lambda^{2}P^{\text{CS}}_{\text{t}}P^{\text{RS}}_{\text{t}}k^{2}_{\text{ch}} + \left| h^{\text{RS}}_{\text{hu},i} \right|^{2}k^{2}_{\text{hu},i}P^{\text{RS}}_{\text{t}}$, and $\mathscr{C}_{\text{a}} = \sum_{i\neq1}^{L}\alpha_{i}\left( \Lambda^{2}P^{\text{RS}}_{\text{t}}P^{\text{CS}}_{\text{t}}\left| h^{\text{RS}}_{\text{ch}} \right|^{2}\left| h^{\text{RS}}_{\text{hu},i} \right|^{2} \right)$.

For a fair comparison between the HAPS-RS and HAPS-RIS scenarios, the total transmit power is conserved. Specifically, the CS transmit power in the HAPS-RIS case ($P^{\text{CS}}_{\text{t}}$) is set equal to the sum of the CS and RS transmit powers in the HAPS-RS case, i.e., $\bar{P}^{\text{CS}}_{\text{t}}=\beta P^{\text{CS}}_{\text{t}}$ and $P^{\text{RS}}_{\text{t}}=(1-\beta)P^{\text{CS}}_{\text{t}}$, where $0<\beta <1$. The optimal value of $\beta$ can be obtained  via a one-dimensional numerical search that maximizes the sum-rate performance, as commonly adopted in the literature~\cite{alfattani2022multi}.


\color{black}Finally, considering half-duplex relaying, the sum rate of the system is computed as \color{black}

\begin{equation}
    C_{\text{RS},s_{i}} = \sum_{i=1}^{L}\frac{1}{2}\text{log}_{2}\left( 1+\gamma_{\text{RS},s_{i}} \right).
\end{equation}

\subsection{HAPS-RIS-assisted Communication}
As depicted in Fig.~1, the CS employs superposition coding and allocates transmission power according to the ordered effective cascaded channel gains, i.e., $\left| {\textbf{h}}^{\text{RIS}}_{\text{hu,1}}{}^{H}\psi_{\text{RIS}}\textbf{H}^{\text{RIS}}_{\text{ch}} \right| < \left| {\textbf{h}}^{\text{RIS}}_{\text{hu,2}}{}^{H}\psi_{\text{RIS}}\textbf{H}^{\text{RIS}}_{\text{ch}} \right| < ... < \left| {\textbf{h}}^{\text{RIS}}_{\text{hu},i}{}^{H}\psi_{\text{RIS}}\textbf{H}^{\text{RIS}}_{\text{ch}} \right|$. The transmitted signal is reflected by the HAPS-RIS toward the ground users. Accordingly, the received signal at the $i$-th user is given by
\begin{equation}
    y_{\text{RIS},i} = {\textbf{h}}^{\text{RIS}}_{\text{hu},i}{}^{H}\psi_{\text{RIS}}\textbf{H}^{\text{RIS}}_{\text{ch}}\left( \sqrt{P^{\text{CS}}_{\text{t}}}x_{\text{sc}} + \eta^{\text{RIS}}_{\text{t,ch}} \right) + \eta^{\text{RIS}}_{\text{r,hu},i} + \aleph,
\end{equation}
where $\psi_{\text{RIS}}=\text{diag}\left( e^{j\theta_{1}}, e^{j\theta_{2}},..., e^{j\theta_{N}} \right)$ denotes the RIS phase-shift matrix, and $\theta_{n}$ represents the phase shift of the $n$-th reflecting element.

The channel between the CS and the HAPS-RIS is modeled as a Rician fading channel, denoted by $\textbf{H}^{\text{RIS}}_{\text{ch}} \in\mathbb{C}^{N\times1}$, as
\begin{equation}\textbf{H}^{\text{RIS}}_{\text{ch}} = \sqrt{\frac{G_{\text{CS}}}{\zeta_{\text{ch}}}\frac{Z_{\text{ch}}}{Z_{\text{ch}}+1}}\bar{\textbf{H}}^{\text{RIS}}_{\text{ch}} + \sqrt{\frac{G_{\text{CS}}}{\zeta_{\text{ch}}}\frac{1}{Z_{\text{ch}}+1}}\tilde{\textbf{H}}^{\text{RIS}}_{\text{ch}}.
\end{equation}
The channel between the HAPS-RIS and the $i$-th user is defined as ${\textbf{h}}^{\text{RIS}}_{\text{hu},i} \in\mathbb{C}^{N_i\times1}$, where $N_i$ denotes the number of RIS elements allocated to user $i$, satisfying $N = \sum_{i=1}^{L} N_i$. It is modeled as

\begin{equation}
    {\textbf{h}}^{\text{RIS}}_{\text{hu},i} = \sqrt{\frac{G_{\text{U}}}{\zeta_{\text{hu},i}}\frac{Z_{\text{hu},i}}{Z_{\text{hu},i}+1}}\bar{\textbf{h}}^{\text{RIS}}_{\text{hu},i} + \sqrt{\frac{G_{\text{U}}}{\zeta_{\text{hu},i}}\frac{1}{Z_{\text{hu},i}+1}}\tilde{\textbf{h}}^{\text{RIS}}_{\text{hu},i},
\end{equation}
where $\bar{\textbf{H}}^{\text{RIS}}_{\text{ch}} \in\mathbb{C}^{N\times1}$ and 
$\bar{\textbf{h}}^{\text{RIS}}_{\text{hu},i} \in\mathbb{C}^{N_i\times1}$ denote the LoS components, 
while $\tilde{\textbf{H}}^{\text{RIS}}_{\text{ch}} \in\mathbb{C}^{N\times1}$ and 
$\tilde{\textbf{h}}^{\text{RIS}}_{\text{hu},i} \in\mathbb{C}^{N_i\times1}$ represent the NLoS components whose elements follow $\mathcal{CN}(0,1)$.

The distortion noise terms $\eta^{\text{RIS}}_{\text{t,ch}}$ and $\eta^{\text{RIS}}_{\text{r,hu},i}$ are modeled as independent Gaussian random variables, where 
$\eta^{\text{RIS}}_{\text{t,ch}} \sim \mathcal{CN}(0,k^{2}_{\text{t,ch}}P^{\text{CS}}_{\text{t}})$ 
and 
$\eta^{\text{RIS}}_{\text{r,hu},i} \sim \mathcal{CN}(0,k^{2}_{\text{r,hu},i}P^{\text{CS}}_{\text{t}}\left| {\textbf{h}}^{\text{RIS}}_{\text{hu},i}{}^{H}\psi_{\text{RIS}} \right|^{2})$.

\color{black}Due to the passive nature of RIS elements, the individual CS--HAPS-RIS and HAPS-RIS--user channels cannot be estimated separately. Instead, the cascaded end-to-end channel is estimated. \color{black}Accordingly, the effective cascaded channel for the $i$-th user is defined as
$\textbf{g}^{\text{RIS}}_{i} = {\textbf{h}}^{\text{RIS}}_{\text{hu},i}{}^{H}\psi_{\text{RIS}} \textbf{H}^{\text{RIS}}_{\text{ch}}$. The cascaded channel is assumed to be imperfectly known and modeled as
$\textbf{g}^{\text{RIS}}_{i} = \hat{\textbf{g}}^{\text{RIS}}_{i} + \Delta \textbf{g}^{\text{RIS}}_{i}$,
where $\hat{\textbf{g}}^{\text{RIS}}_{i}$ denotes the estimated cascaded channel, and 
$\Delta \textbf{g}^{\text{RIS}}_{i}$ represents the corresponding estimation error. 
The estimation error is modeled as a complex Gaussian random variable with zero mean and variance proportional to the channel power, i.e., $\Delta \textbf{g}^{\text{RIS}}_{i} \sim \mathcal{CN}\left(0, \epsilon \, \mathbb{E}\left[\left|\textbf{g}^{\text{RIS}}_{i}\right|^{2}\right]\right)$, where $\epsilon$ denotes the level of CSI imperfection.

Considering the presence of HWI and imperfect CSI, the SINR at the $i$-th user is given by

\begin{equation}
    \gamma_{\text{RIS},s_{i}}= \frac{\alpha_{1}\left| \mathbf{g}^{\text{RIS}}_{i} \right|^{2}P^{\text{CS}}_{\text{t}}}{\mathscr{C}_{\text{b}}+\Delta_{\text{b}}+\sigma^{2}},
\end{equation}
where 
$\Delta_{\text{b}}=\left| \mathbf{g}^{\text{RIS}}_{i} \right|^{2}P^{\text{CS}}_{\text{t}}k^{2}_{\text{ch}}$. 
Additionally, $\mathscr{C}_{\text{b}} = \sum_{j\neq i}^{L}\alpha_{j}\left( \left| \mathbf{g}^{\text{RIS}}_{j} \right|^{2}P^{\text{CS}}_{\text{t}} \right)$. Finally, the achievable sum rate is expressed as

\begin{equation}
    C_{\text{RIS},s_{i}}= \sum_{i=1}^{L}\log_{2}\left( 1+\gamma_{\text{RIS},s_{i}} \right).
\end{equation}

\subsection{Path Loss Model}

The path loss between the CS and the HAPS, as well as between the HAPS and the ground users, is modeled based on the Friis transmission equation. Let $\left( x^{\text{CS}}, y^{\text{CS}}, z^{\text{CS}} \right)$, $\left( x_{i}, y_{i}, z_{i} \right)$, $\left( x^{\text{HAPS}}, y^{\text{HAPS}}, z^{\text{HAPS}} \right)$, $\left( x_{m}^{\text{RIS}}, y_{m}^{\text{RIS}}, z_{m}^{\text{RIS}} \right)$, and $\left( x^{\text{RS}}, y^{\text{RS}}, z^{\text{RS}} \right)$ denote the spatial coordinates of the CS, the $i$-th ground user, the HAPS, the $m$-th RIS element, and the RS, respectively. Accordingly, the large-scale path loss between the CS and the HAPS is given by

\begin{equation}
\begin{array}{l}
\zeta_{\text{ch}} =
\left( \frac{4\pi f_{c}}{c} \right)^{2} \cdot \\[3pt]
\left(\left( x^{\text{HAPS}}-x^{\text{CS}} \right)^{2}
+ \left( y^{\text{HAPS}}-y^{\text{CS}} \right)^{2}
+ \left( z^{\text{HAPS}}-z^{\text{CS}} \right)^{2}\right).
\end{array}
\end{equation}

Similarly, the path loss between the $m$-th RIS element mounted on the HAPS and ground user $i$ is expressed as
\begin{equation}
\begin{array}{l}
\zeta_{\text{hu},i} =
\left( \frac{4\pi f_\text{c}}{c} \right)^{2} \cdot \\[3pt]
\left(\left( x_{i}-x_{m}^{\text{RIS}} \right)^{2}
+ \left( y_{i}-y_{m}^{\text{RIS}} \right)^{2}
+ \left( z_{i}-z_{m}^{\text{RIS}} \right)^{2}\right).
\end{array}
\end{equation}
Here, $f_\text{c}$ denotes the carrier frequency, while $c$ represents the speed of light. In this geometric configuration, the physical size of the RIS is negligible compared to the HAPS altitude and has a limited impact on system performance~\cite{azizi2025exploring}. Therefore, without loss of generality, the positions of the RIS elements are approximated as follows (note that the RIS dimensions are $5$ m $\times$ $6$ m, whereas the CS--HAPS and HAPS--ground user distances exceed 20 km.): $
\left( x^{\text{RIS}}_m,y^{\text{RIS}}_m,z^{\text{RIS}}_m \right) 
\approx 
\left( x^{\text{HAPS}},y^{\text{HAPS}},z^{\text{HAPS}} \right), 
\quad \forall m$.
Similarly, the location of the RS is approximated as $
\left( x^{\text{RS}},y^{\text{RS}},z^{\text{RS}} \right) 
\approx 
\left( x^{\text{HAPS}},y^{\text{HAPS}},z^{\text{HAPS}} \right)
$.



\section{Energy Efficiency Analysis}

The energy efficiency of the HAPS-RS- and HAPS-RIS-assisted communication systems is defined as the ratio of the achievable sum rate to the total power consumption. Accordingly, the EE can be expressed as \cite{khennoufa2025multi}

\begin{equation}
E_{q} = \frac{C_{q,s_i}}{P_{q}},
\end{equation}
where \( q \in \{\text{RS, RIS}\} \), and $P_q$ denotes the total power consumption of the corresponding system configuration. The total power consumption is modeled as
\begin{equation}
P_{q}=P^{\text{CS}}_{\text{t}}+\Gamma_q+\sum_{i=1}^{L}P_{\text{u},i}, 
\end{equation}
where $\Gamma_{\text{RS}} = P^{\text{RS}}_{\text{t}} + P_{\text{RS}}$ and $\Gamma_{\text{RIS}} = NP_{\text{sw}} + NP_{\text{dc}}$. Here, $P^{\text{CS}}_{\text{t}}$ represents the transmit power of the CS. For the RIS-assisted system, $P_{\text{sw}}$ and $P_{\text{dc}}$ denote the power consumption associated with the phase-shifting switches and the DC biasing circuits of each RIS element, respectively. For the RS-assisted system, $P^{\text{RS}}_{\text{t}}$ denotes the transmit power of the RS, while $P_{\text{RS}}$ accounts for its payload-related power consumption~\cite{alfattani2022multi}. In addition, $P_{\text{u},i}$ represents the power consumption of the $i$-th ground user.

\section{Numerical Results and Discussion}

In this section, the performance of the HAPS-RS- and HAPS-RIS-assisted communication systems is evaluated through Monte Carlo simulations with $10^5$ independent realizations. Also, a two-user NOMA scenario ($L=2$) is considered, as increasing the number of users introduces additional complexity in successive interference cancellation (SIC) and power allocation. The CS is located at $(-5 \ \text{km}, -5 \ \text{km}, 0 \ \text{km})$, while the HAPS is positioned at $(0 \ \text{km}, 0 \ \text{km}, 20 \ \text{km})$. The ground users are located at $(15 \ \text{km}, 1 \ \text{km}, 0 \ \text{km})$ for $U_1$ and $(0 \ \text{km}, 1 \ \text{km}, 0 \ \text{km})$ for $U_2$. For the HAPS-RS configuration, 
the payload power consumption is assumed to be up to 1 kW, consistent with the specifications of the X-Station HAPS platform developed by StratXX~\cite{alfattani2022multi}. The remaining simulation parameters are summarized in Table~I.

\begin{table}
\fontsize{8.0pt}{8.0pt}\selectfont
\begin{center}
\caption{Simulation Parameters.}
\renewcommand{\arraystretch}{1} 
\begin{tabular}[!h]{|c| c||c| c|}
    \hline
    \textbf{Parameter} & \textbf{Value} & \textbf{Parameter} & \textbf{Value}\\
    \hline
    $L$ & 2 & $k = K^{2}_{\text{ch}}=K^{2}_{\text{hu},i}$ & $0.1$  \\
    \hline
    $\alpha_1$ & $0.8$ & $\varepsilon$ & $0.1$  \\
    \hline
    $\alpha_2$ & $0.2$ & $P_{\text{sw}}$ &  $7.8$ mW \cite{karaman2025trade} \\
    \hline
    $f_{\text{c}}$ & $2$ GHz  & $P_{\text{dc}}$ & $-5$ dBm \cite{khennoufa2025multi}  \\
    \hline
    $G_{\text{CS}}$ & $43.2$ dB  & $P_{\text{RS}}$ & $1$ kW \cite{alfattani2022multi} \\
    \hline
    $G_{\text{RS}}$ & $15$ dB \cite{alfattani2022multi} & $P_{\text{u},i}$ & $10$ dBm \\
    \hline
    $G_{\text{U}}$ & $0$ dB   & $N_\text{F}$ & $7$ dB \\
    \hline
    $Z_{\text{ch}}=Z_{\text{hu},i}$  & $15$ dB & $\sigma^2$ & $-107$ dBm  \\
        
    \hline
\end{tabular}
\end{center}
\end{table}

Fig.~2 illustrates the impact of the total number of RIS elements on the achievable sum rate for both HAPS-RIS and HAPS-RS systems under OMA and NOMA schemes. Here, an equal number of RIS elements is allocated to both users. It is observed that the HAPS-RIS configuration significantly benefits from increasing the number of reflecting elements, particularly beyond $N \geq 15 \times 10^3$, where a notable improvement in sum-rate performance is achieved under ideal conditions. Furthermore, NOMA consistently outperforms OMA due to its ability to exploit power-domain multiplexing. However, the presence of HWI and imperfect CSI introduces a considerable performance degradation across all configurations. The superior performance of HAPS-RIS over HAPS-RS under non-ideal conditions can be attributed to the passive nature of RIS, which avoids noise amplification. In contrast, the RS actively forwards the received signal, thereby amplifying not only the desired signal but also hardware-induced distortions and channel estimation errors. \color{black} Furthermore, in the HAPS-RIS system, the sum-rate exhibits saturation beyond $10 \times10^3$ elements due to the impact of imperfect CSI. \color{black} In particular, the combined effect of HWI and CSI results in the most pronounced performance loss, highlighting the importance of accounting for practical system limitations.

\begin{figure}[!t]
\centering
\includegraphics[width=2.6in]{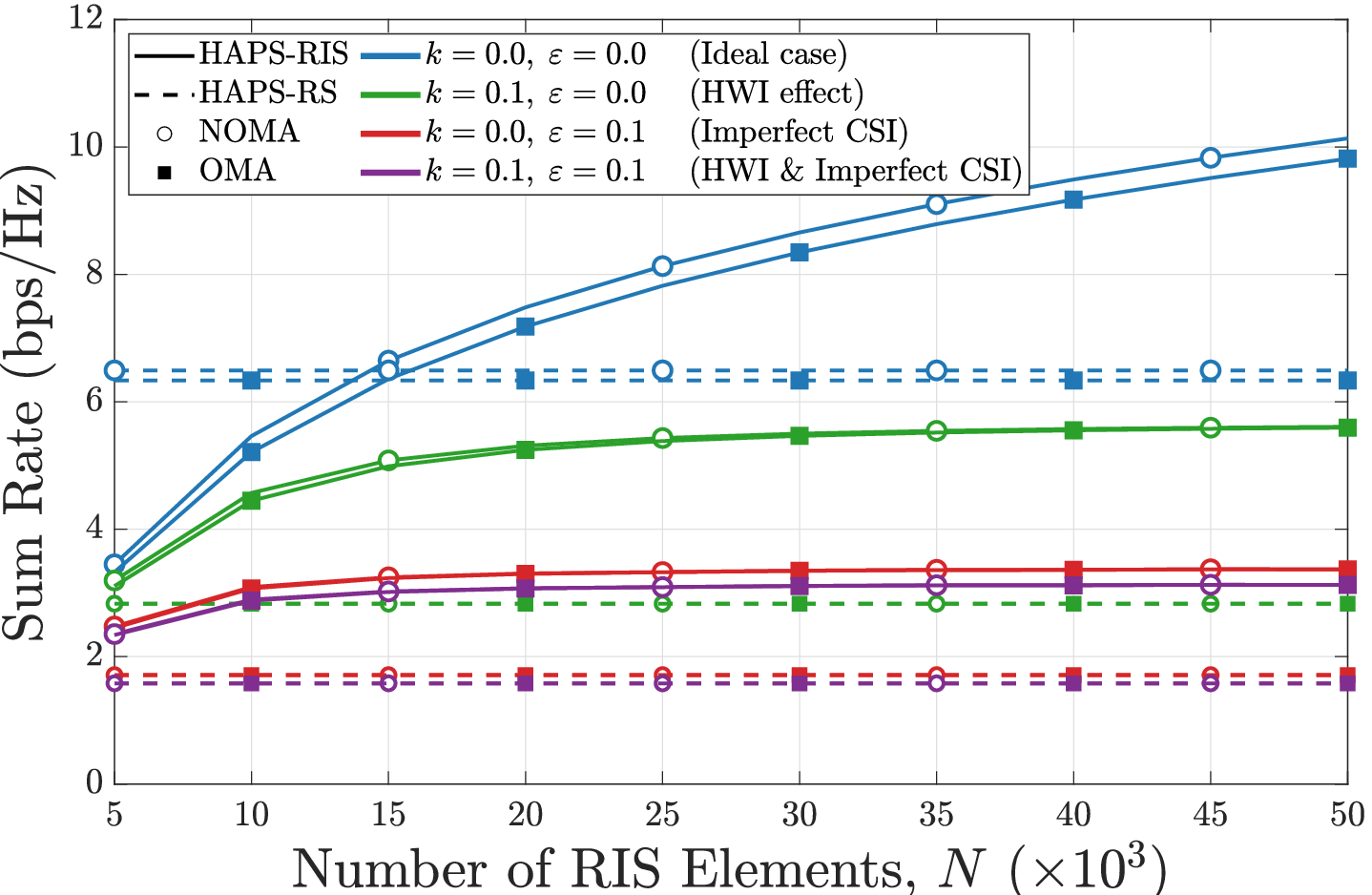}
\caption{Sum-rate vs. the number of RIS elements for HAPS-RIS and HAPS-RS systems under impairments ($P^{\text{CS}}_{\text{t}}=40$ dBm).}
\end{figure}

The energy efficiency as a function of the number of RIS elements for both HAPS-RIS and HAPS-RS systems is depicted in Fig.~3. It is observed that, unlike the sum-rate behavior, the energy efficiency of HAPS-RIS decreases as the number of RIS elements increases. This is mainly due to the linear growth in power consumption associated with the RIS hardware, while the corresponding sum-rate improvement becomes marginal beyond a certain point. In contrast, the HAPS-RS configuration exhibits relatively constant energy efficiency, as it does not depend on the number of RIS elements. Similar to Fig.~2, HWI and imperfect CSI significantly degrade performance, with their combined effect resulting in the lowest energy efficiency. These highlight a key trade-off between spectral and energy efficiency in HAPS-RIS systems.
\begin{figure}[!t]
\centering
\includegraphics[width=2.6in]{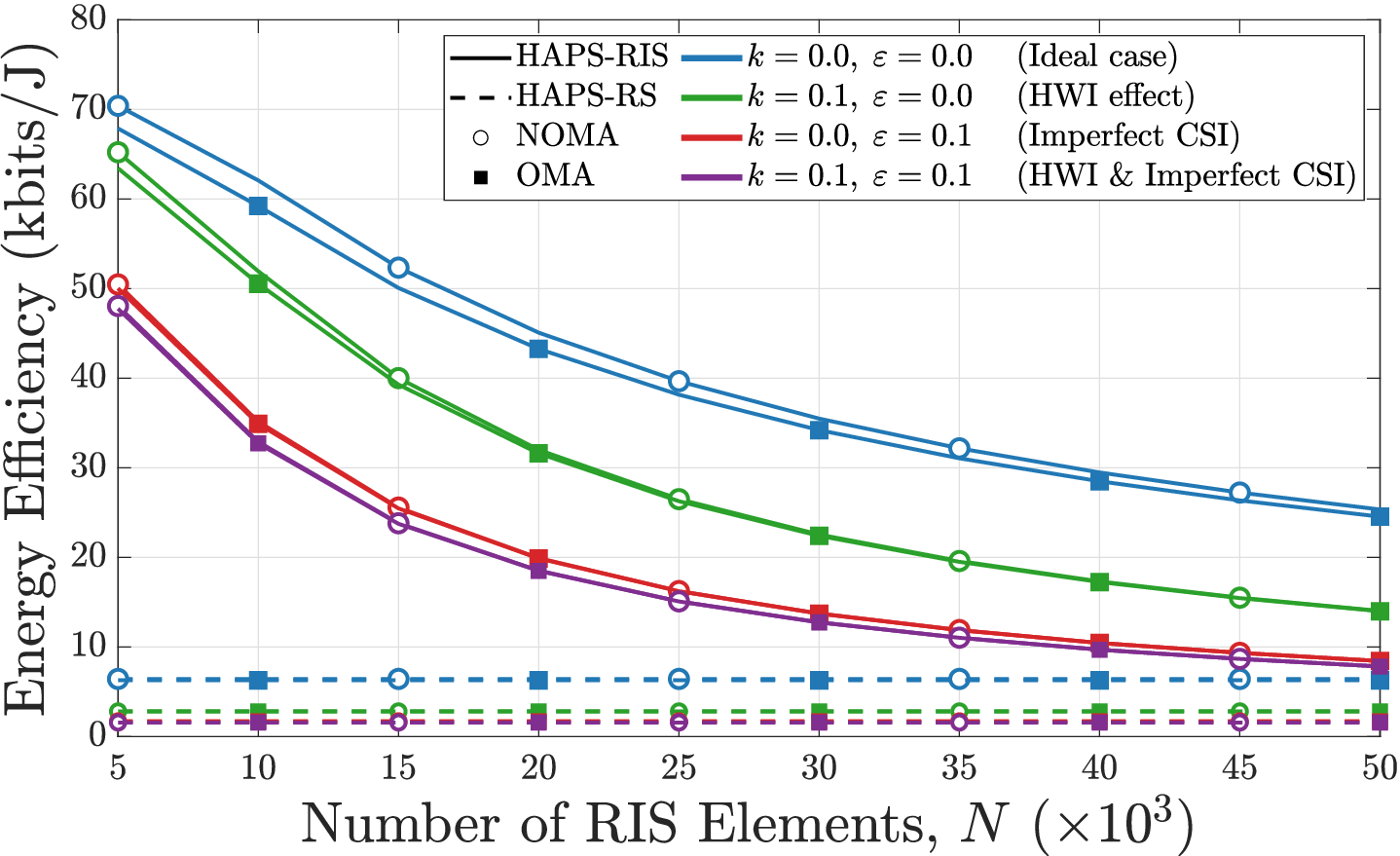}
\caption{Energy eff. vs. the number of RIS elements under impairments ($P^{\text{CS}}_{\text{t}}= 40$ dBm).}
\end{figure}

Fig.~4 illustrates the sum-rate performance as a function of the transmit power, where the number of RIS elements is fixed at $N = 30 \times 10^3$ for the HAPS-RIS scenario. Here, an equal number of RIS elements is allocated to both users. It is observed that the sum rate increases with transmit power for all configurations. However, the presence of HWI and imperfect CSI significantly limits the achievable gains, particularly at high transmit power levels, where performance saturation is observed. This behavior indicates that HAPS-RS is advantageous in the low-power regime due to active amplification, while HAPS-RIS becomes superior at high transmit power levels by mitigating noise and distortion amplification.

\begin{figure}[!t]
\centering
\includegraphics[width=2.6in]{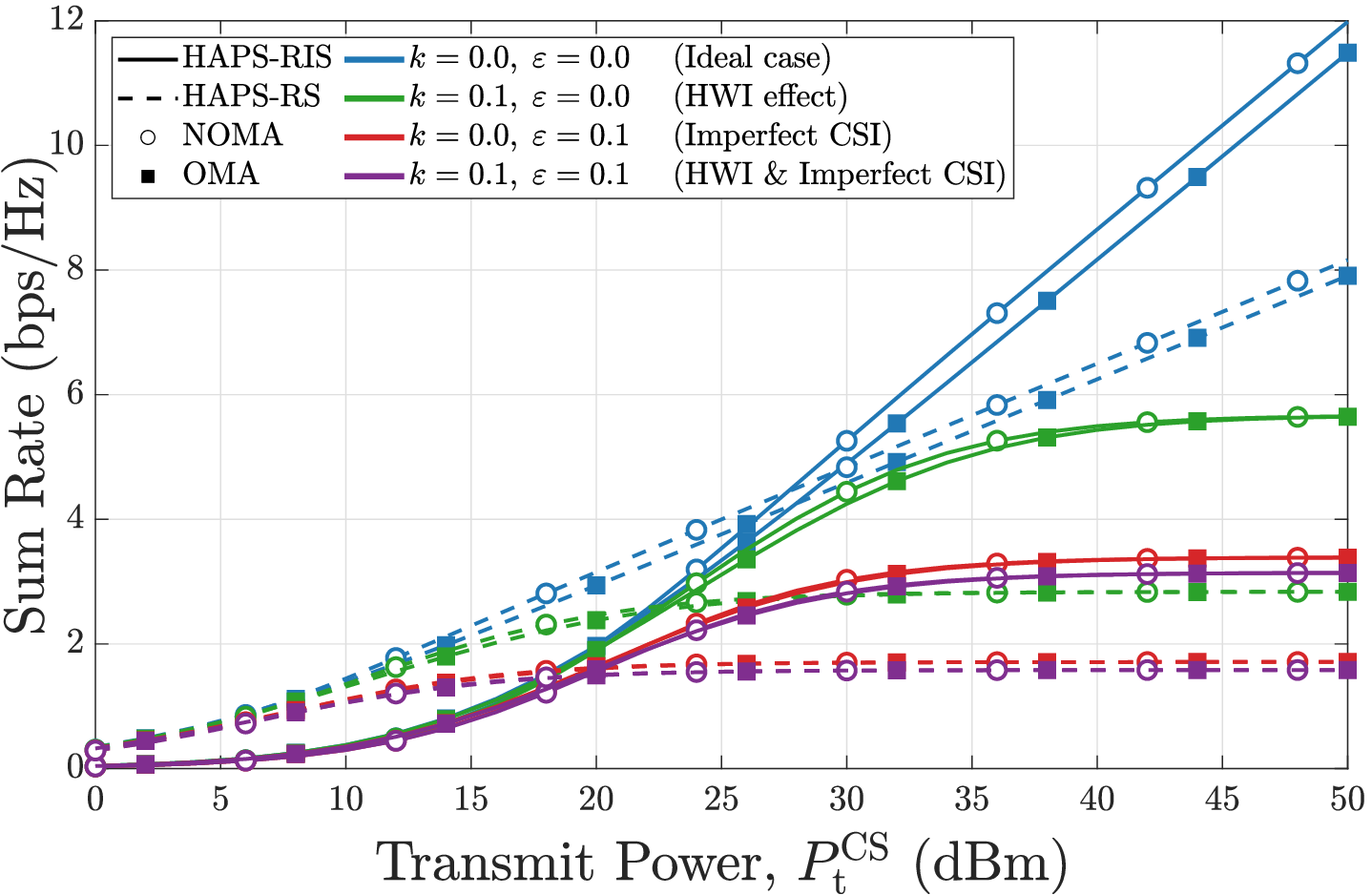}
\caption{Sum-rate vs. transmit power under  impairments ($N=30 \times 10^3$).}
\end{figure}

The energy efficiency as a function of the transmit power is presented in Fig.~5. In the HAPS-RIS scheme, it is observed that the energy efficiency initially increases, reaches a peak, and then decreases due to the dominance of power consumption over achievable rate gains. HAPS-RIS significantly outperforms HAPS-RS, particularly in the moderate-to-high transmit power regime, while the inferior performance of HAPS-RS is mainly attributed to its high power consumption due to active relaying. Moreover, HWI and imperfect CSI reduce the peak energy efficiency and shift the optimal operating point. These results confirm the existence of an optimal transmit power that maximizes energy efficiency in practical HAPS-assisted communication systems.

\begin{figure}[!t]
\centering
\includegraphics[width=2.6in]{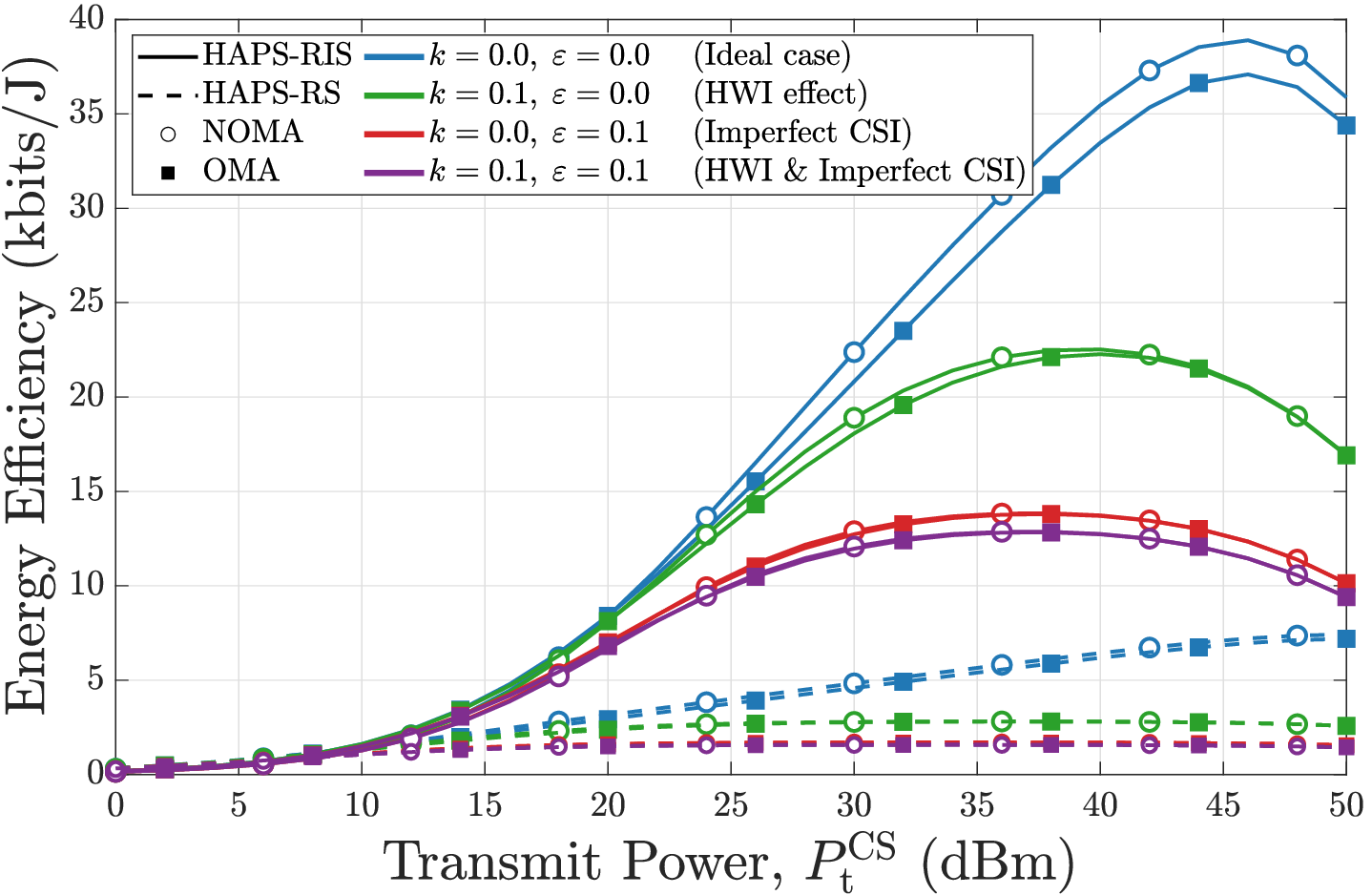}
\caption{Energy eff. vs. transmit power under impairments ($N=30 \times 10^3$).}
\end{figure}

To investigate the impact of user distance on NOMA performance, Fig.~6 illustrates the sum-rate behavior as a function of the far user position, while the location of the near user $U_2$ is kept fixed. In this scenario, the transmit power and the total number of RIS elements are set to $P^{\text{CS}}_{\text{t}}=40$ dBm and $N=10 \times 10^3$, respectively, and the RIS elements are equally allocated between users. We observe that the overall sum rate decreases as the far user moves away due to increased path loss. However, the NOMA gain increases with the distance between users, as stronger channel disparity is created. In HAPS-RIS systems with dominant LoS conditions, users tend to experience similar channel gains when located close to each other, which limits the effectiveness of NOMA. As the distance between users increases, this limitation is alleviated, allowing NOMA to better exploit power-domain multiplexing. This result highlights that user spatial separation plays a critical role in unlocking the potential of NOMA in HAPS-RIS systems.

\begin{figure}[!t]
\centering
\includegraphics[width=2.35in]{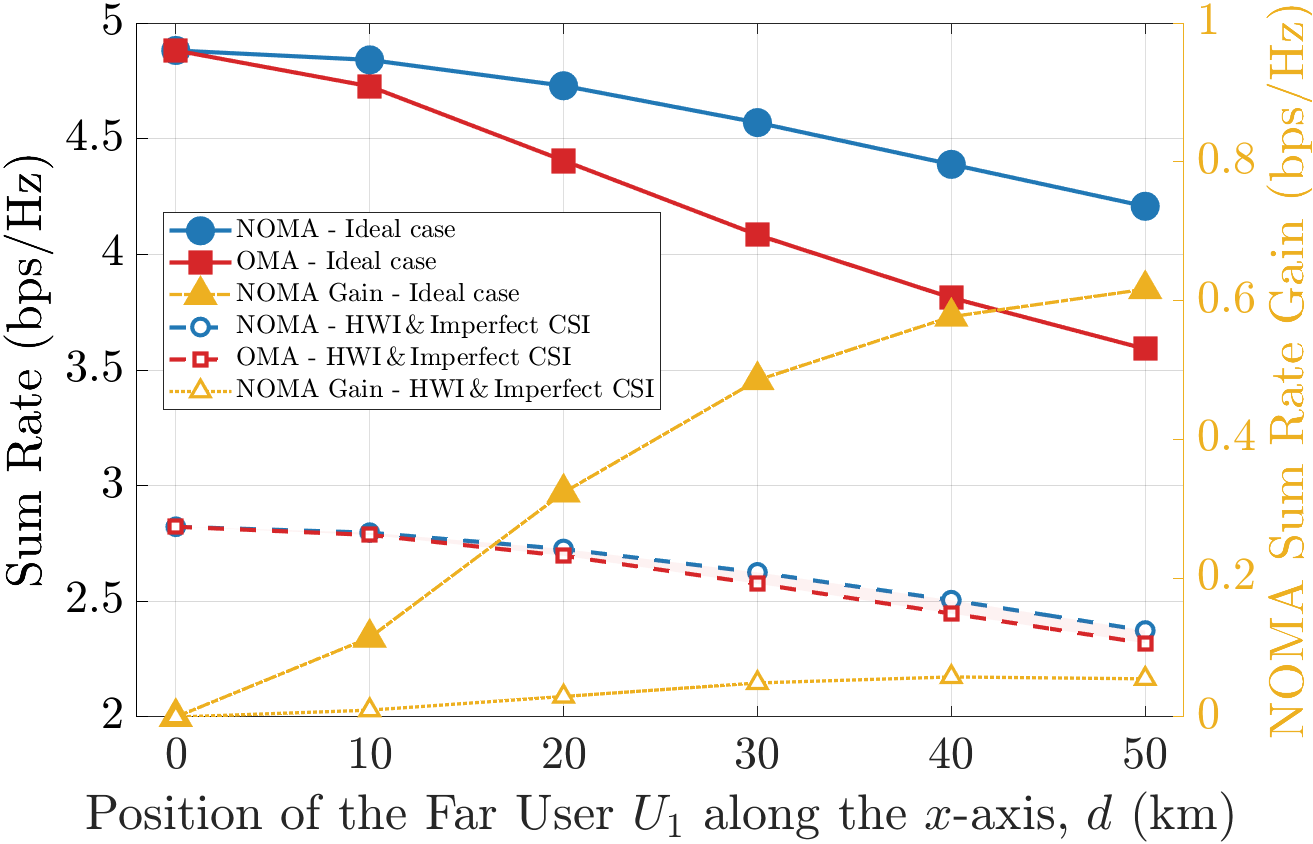}
\caption{Impact of far user location on sum-rate and NOMA gain ($N=10 \times 10^3$, $P^{\text{CS}}_{\text{t}} = 40$ dBm, $k=0.1, \epsilon=0.1$).}
\end{figure}

Fig.~7 investigates the impact of RIS elements allocation on the system performance. The total number of RIS elements is fixed at $N = 10 \times 10^3$, with $P^{\text{CS}}_{\text{t}} = 40$ dBm. The ground users are located at $(10 \ \text{km}, 1 \ \text{km}, 0 \ \text{km})$ for $U_1$ and $(0 \ \text{km}, 1 \ \text{km}, 0 \ \text{km})$ for $U_2$. It is observed that allocating more RIS elements to the far user improves the overall sum-rate performance by increasing the channel gain disparity between users. In contrast, equal allocation leads to a moderate performance level, as it fails to fully exploit the potential of power-domain multiplexing. \color{black} Also, under HWI and imperfect CSI, the optimal allocation point shifts from $2500$ to $4500$ elements, indicating that the impairment level directly influences the RIS element allocation strategy. \color{black} This demonstrates that optimum RIS elements allocation can effectively emulate channel heterogeneity, which is essential for maximizing NOMA performance in LoS-dominant HAPS-RIS systems.

\begin{figure}[!t]
\centering
\includegraphics[width=2.61in]{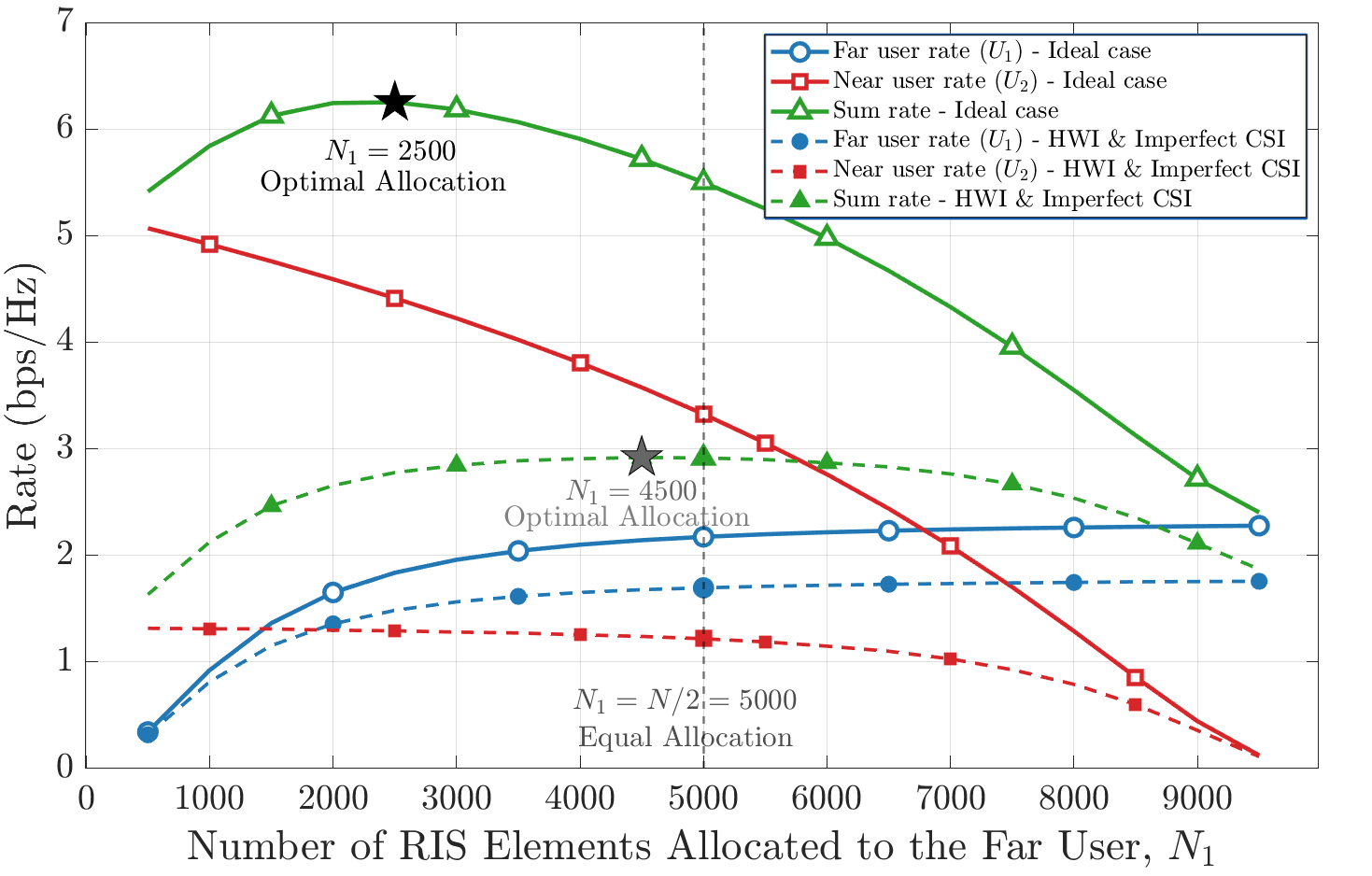}
\caption{Impact of RIS elements allocation on the sum-rate performance for HAPS-RIS systems ($N=10 \times 10^3$, $P^{\text{CS}}_{\text{t}} = 40$ dBm, $k=0.1, \epsilon=0.1$).}
\end{figure}

Assuming a unit-cell size of $(0.2\lambda)^2$ at a carrier frequency of $2$ GHz~\cite{safwanGlobecom}, the considered RIS configurations with $5 \times 10^3$ to $30 \times 10^3$ elements correspond to a total surface area ranging from approximately $4.5~\mathrm{m}^2$ to $27~\mathrm{m}^2$. From a power consumption perspective, by assuming $P_{\text{RIS}}=7.8$ mW per element~\cite{safwanGlobecom}, the total power requirement varies between $39~\mathrm{W}$ and $234~\mathrm{W}$. This level of power consumption is significantly lower than that of conventional BS or RS systems~\cite{alfattani2022multi}. Moreover, such power levels can be sustained by $1 \ \text{m}^{2}$ PV panels with a relatively small footprint under favorable conditions~\cite{sayed2024feasibility}. This highlights the strong energy efficiency advantage of HAPS-RIS systems, making them a highly promising solution for sustainable and energy-constrained 6G NTN.

\section{Conclusion}

This paper investigates the performance of HAPS-RS- and HAPS-RIS-assisted communication systems under OMA and NOMA schemes by considering practical impairments such as HWI and imperfect CSI. The results demonstrated that HAPS-RIS achieves superior sum-rate and energy efficiency compared to HAPS-RS under non-ideal conditions, owing to its passive structure that avoids noise and distortion amplification. It was further shown that RIS elements allocation and user spatial distribution play a critical role in enhancing NOMA performance, particularly in LoS-dominant environments, where appropriate allocation can effectively create channel disparity. 
These findings highlight the importance of impairment-aware system design and resource allocation in HAPS-assisted networks. Future work may focus on adaptive RIS allocation and dynamic user pairing strategies.


\bibliographystyle{IEEEtran}
\bibliography{bibliography}

\vfill

\end{document}